\documentclass[a4paper,11pt]{article}
\usepackage{jheppub}
\usepackage[utf8x]{inputenc}
\usepackage{graphicx}
\usepackage{float}
\usepackage[T1]{fontenc}
\usepackage{epsfig}
\usepackage{color}
\usepackage{tikz}
\usepackage{amsmath,mathtools}
\usepackage{subfloat}
\usepackage{amsfonts}
\usepackage{braket}
\usepackage{cleveref}
\usepackage{physics}
\usepackage{epstopdf}
\usepackage{caption}
\usepackage{subcaption}
\usepackage{enumitem}
\usepackage{etoolbox}
\usepackage{orcidlink}
\usepackage{comment}
\graphicspath{}
\usepackage[titletoc,toc,title]{appendix}
\usepackage{hyperref}
\hypersetup{
	colorlinks=true,
	linkcolor=blue,
	filecolor=red,      
	urlcolor=blue,
	citecolor=red
}

\newcommand{\cI}{c_{\rm I}}
\newcommand{\cII}{c_{\rm II}}
\newcommand{\LI}{L_{\rm I}}
\newcommand{\LII}{L_{\rm II}}
\newcommand{\rhoI}{\rho_{\rm I}^0}
\newcommand{\rhoII}{\rho_{\rm II}^0}

\title{Defects extremal surfaces and interface CFTs}
\author[]{Ankur Dey\orcidlink{0009-0001-1077-0442}\;}

\affiliation[]{Department of Physics,\\Indian Institute of Technology Kanpur,\\208016, India}
\emailAdd{ankurd21@iitk.ac.in}

\date{}

\abstract{

We propose a generalization of the defect extremal surface (DES) prescription to holographic interface conformal field theories (ICFTs), extending its applicability beyond the AdS$_3$/BCFT$_2$ framework. We consider a holographic ICFT$_2$ comprising two CFT$_2$s coupled across an interface, dual to two asymptotically AdS$_3$ geometries joined by a codimension-one end-of-the-world (EOW) brane. In the corresponding effective lower-dimensional description obtained via partial dimensional reduction on the brane-boundary combination, the theory on the brane is coupled to gravity, thereby enabling the computation of fine-grained entanglement entropy via the island formula. We demonstrate that the generalized DES prescription exactly reproduces the effective lower-dimensional results for finite subsystems in time-independent scenarios in the large brane tension limit, corresponding to weak gravitational coupling on the EOW brane. We further compute the entanglement entropy of semi-infinite bipartite pure states in radiation subsystems coupled to $2d$ eternal black holes, and obtain the corresponding Page curves.

}

\begin{document}
\maketitle


\section{Introduction}\label{sec_Intro}

For the past few decades, the black hole information loss paradox has attracted considerable research attention in the context of formulating a consistent quantum theory of gravity. More recently, a possible resolution of this paradox has been proposed in toy models of quantum field theories coupled to semiclassical gravity, which successfully reproduce the Page curve \cite{Page:1993wv, Page:1993df, Page:2013dx} for Hawking radiation. This development involved the emergence of the island formula for the fine-grained entanglement entropy of Hawking radiation, motivated by the quantum extremal surface prescription introduced in \cite{Engelhardt:2014gca}. Within this framework, spacetime regions referred to as islands appear at late times in the entanglement wedge of subsystems in the radiation bath, leading to the purification of the outgoing Hawking quanta \cite{Almheiri:2019hni, Almheiri:2019psf, Almheiri:2019qdq, Almheiri:2019psy, Almheiri:2019yqk, Almheiri:2020cfm}.

A higher-dimensional perspective on the island formulation is provided by double holography \cite{Almheiri:2019hni, Sully:2020pza, Rozali:2019day, Chen:2020uac, Chen:2020hmv, Grimaldi:2022suv, Suzuki:2022xwv, Geng:2020qvw, Geng:2020fxl, Geng:2021iyq, Geng:2021mic, Geng:2021hlu}, where a $(d+1)$-dimensional gravitational braneworld theory is holographically dual to a $d$-dimensional conformal field theory (CFT$_d$) coupled to semiclassical gravity on the end-of-the-world (EOW) brane.\footnote{Since the brane description may be obtained via a partial reduction of the bulk AdS$_3$ geometry with the EOW brane, it is also referred to as the lower-dimensional effective description \cite{Deng:2020ent, Chu:2021gdb, Li:2021dmf, Basu:2022reu, Shao:2022wrm, Basu:2024xjq}.} The gravity-matter theory on the brane is, in turn, dual to a quantum mechanical system residing at the boundary of the CFT$_d$. Within this framework, the island prescription is precisely equivalent to the Ryu-Takayanagi (RT) prescription \cite{Ryu:2006bv, Ryu:2006ef} in the bulk AdS$_{d+1}$ geometry. Closely related to this construction is the AdS/BCFT correspondence \cite{Takayanagi:2011zk, Fujita:2011fp, Sully:2020pza, Kastikainen:2021ybu, Izumi:2022opi, Cavalcanti:2018pta, Magan:2014dwa, Cavalcanti:2020rsp, Takayanagi:2020njm}, which realizes the holographic dual of a boundary conformal field theory (BCFT) \cite{Cardy:2004hm} as an asymptotically AdS geometry truncated by a codimension-one EOW brane with Neumann boundary conditions. In this framework, the homology constraint is modified such that the RT surfaces may terminate on the EOW brane. For recent developments in double holography, see \cite{Liu:2023ggg,Ling:2021vxe,Akal:2020wfl,Miao:2020oey,Myers:2024zhb}.

Building on the AdS/BCFT framework described above, the authors in \cite{Deng:2020ent} proposed the defect extremal surface (DES) prescription as the holographic counterpart of the island formula in $d=2$ dimensions. In this construction, the bulk is treated as a defect spacetime with conformal matter localized on the EOW brane, and the gravitational region relevant for the quantum extremal surface emerges through a partial dimensional reduction of the bulk geometry. A transparent boundary condition applies across the interface between the gravitational and non-gravitational regions of the CFT$_2$. The fine-grained entanglement entropy computed using the DES formula agrees with that obtained through the boundary quantum extremal surface prescription. Subsequently, the DES prescription was employed in \cite{Chu:2021gdb} to compute the entanglement entropy of radiation subsystems coupled to a $2d$ eternal black hole. For further developments involving the DES prescription see \cite{Deng:2023pjs,Basu:2022reu, Shao:2022wrm, Li:2021dmf,Basu:2024xjq,Wang:2021xih,Dey:2025qms,Dey:2025dlj}.

An important open issue is extending the DES prescription beyond the AdS/BCFT framework to more general defect theories, particularly interface conformal field theories (ICFTs) \cite{Karch:2000gx,Bachas:2001vj}. In this context, ICFTs serve as a natural generalization of BCFTs, in which two CFTs are joined along a common conformally invariant interface. The holographic dual of a two-dimensional ICFT consists of two asymptotically AdS$_3$ geometries glued across a common AdS$_2$ brane satisfying the Israel junction conditions \cite{Anous:2022wqh}. The dynamics of the interface are controlled by the ratio of the central charges of the two CFTs. An effective lower-dimensional description is further obtained through a partial Randall-Sundrum reduction of the bulk geometry, inducing JT gravity on the interface brane coupled to two CFT$_2$ baths. In this framework, the entanglement entropy of subsystems comprising both semi-infinite and finite intervals in the two CFTs was computed using bulk geodesics and the island prescription, which agree exactly in the large tension limit \cite{Anous:2022wqh}.

A distinctive feature of the AdS$_3$/ICFT$_2$ framework is the existence of RT surfaces that cross the interface brane and re-enter the original AdS$_3$ geometry. Such configurations were shown to arise from novel replica wormhole saddles and correspond to induced islands in one of the CFT$_2$ \cite{Anous:2022wqh,Afrasiar:2023nir}. Subsequent investigations demonstrated that the entanglement entropy of semi-infinite radiation subsystems in the thermofield double state exhibits the expected Page curve for a $2d$ eternal black hole induced on the AdS$_2$ brane. For recent works in holographic ICFTs see \cite{Afrasiar:2023nir,Anous:2022wqh,Anand:2025cvg,Basak:2023bnc,Clark:2004sb,Gutperle:2015hcv,Liu:2024oxg}. 

As motivated earlier, it is crucial to extend the DES prescription beyond its original AdS/BCFT framework to holographic ICFTs. In this article, we formulate a generalized DES construction for two asymptotically AdS$_3$ geometries joined across an EOW brane. This framework extends the DES framework to encompass interface-specific features (like unequal central charges, induced islands, replica wormholes, and extremal surfaces that traverse the interface) and provides a higher-dimensional geometric realization of the island formula in interface theories. Interestingly, this is achieved without invoking an explicit, lower-dimensional JT gravity action. We further demonstrate that, in the large brane tension limit, this generalized prescription reproduces the effective lower-dimensional results for the entanglement entropy for subsystems in the radiation baths for both time-independent and time-dependent scenarios.

This paper is organized as follows. In \cref{sec_review} we briefly review the basics of entanglement entropy, the AdS$_3$/BCFT$_2$ framework and double holography, the island formula and the bulk defect extremal surface prescription, and the AdS$_3$/ICFT$_2$ framework. Subsequently, we discuss the generalization of the DES prescription to interface CFTs in \cref{sec_desinterface}, and investigate the entanglement entropy for finite subsystems in time-independent scenarios in \cref{sec_TI}. We further extend our analysis to semi-infinite subsystems in time-dependent scenarios in \cref{sec_TD}. Finally, we summarize our results and discuss future directions in \cref{sec_summary}.


\section{Review of earlier literature}\label{sec_review}

In this section, we review the basics of the pure state entanglement measure known as the entanglement entropy, the double holographic AdS$_3$/BCFT$_2$ framework, the island and the defect extremal surface (DES) formula for the fine-grained entanglement entropy, and the interface CFT (ICFT) model.


\subsection{Entanglement Entropy}

Considering a bipartite system $A \cup B$ in a pure state  described by the density matrix $\rho$, the entanglement entropy of $A$ is defined as 
\begin{align}
S_A = - \mathrm{Tr}_A (\rho_A \log \rho_A)  ,
\end{align}
where $\rho_A = \mathrm{Tr}_B \, \rho$ is the reduced density matrix obtained by tracing out the degrees of freedom of the subsystem $B$. The entanglement entropy may also be computed using the R\'enyi entanglement entropy $S_n$ as 
\begin{equation} \label{Renyi-EE-def}
S_A =  \lim _{n \to 1} S_{n}  = \lim _{n \to 1} \frac{1}{1-n} \log \text{Tr} (\rho_A)^n ,
\end{equation}
where $n$ is the replica index. In a CFT$_2$, using the replica construction this trace $\text{Tr} (\rho_A)^n$ may further be expressed in terms of correlation functions of twist and anti-twist fields $\sigma_n$ and $\bar{\sigma}_n$ with conformal weights $h_n = \bar{h}_n = \frac{c}{24}(n-1/n)$ located at the endpoints of the subsystem $A$ \cite{Calabrese:2004eu, Calabrese:2009ez, Calabrese:2009qy}.

From the bulk perspective, the holographic entanglement entropy may be computed using the Ryu-Takayanagi (RT) prescription \cite{Ryu:2006bv, Ryu:2006ef}. For a subsystem $A$ described on the asymptotic boundary of the dual bulk AdS$_3$ geometry, the holographic entanglement entropy is given as
\begin{align}\label{eq_rt}
S_A = \frac{\mathcal{L}}{4 G_N} ,
\end{align}
where $\mathcal{L}$ is the length of the minimal codimension-two bulk surface homologous to the subsystem $A$, which may be explicitly computed as
\begin{align}\label{eq_geol}
\mathcal{L}=\cosh ^{-1} [-\xi_{ji}].
\end{align}
Here $\xi_{ji}$ is related to the inner product between the position vectors $Y_j$ and $Y_i$ of the two endpoints, which in the Poincar\'e coordinates $(x,\tau,z)$ is given as
\begin{align}\label{eq_ip}
\xi_{ji}=Y_j \cdot Y_i=-\frac{(x_j-x_i)^2+(\tau_j-\tau_i)^2+z_j^2+z_i^2}{2 z_i z_i} .
\end{align}


\subsection{AdS$_3$/BCFT$_2$ and double holography}\label{ssec_adsbcft}

We now discuss the AdS/BCFT framework, which extends the standard AdS/CFT duality to conformal field theories defined on a manifold with a boundary \cite{Cardy:2004hm}. In this context, the $2d$ boundary conformal field theory (BCFT$_2$) is described on the asymptotic boundary of a bulk AdS$_3$ geometry $N$ truncated by a codimension-one end-of-the-world (EOW) brane $Q$ with Neumann boundary conditions. The bulk action is given as \cite{Takayanagi:2011zk,Fujita:2011fp}
\begin{align}
I_0=\frac{1}{16 \pi G_N} \int_N \sqrt{-g} (R-2 \Lambda)+\frac{1}{8 \pi G_N} \int_Q \sqrt{-h} (K-T),
\end{align}
where $g_{\mu \nu}$ and $R$ are the bulk metric and the scalar curvature of the bulk geometry, respectively, while $h_{ab}$ and $K$ are those induced on the EOW brane. $T$ here is the tension on the brane. Variation of the above action with respect to $h_{ab}$ and imposing the Neumann boundary condition we obtain the relation
\begin{align}\label{eq_kab}
K_{ab}=(K-T)h_{ab}.
\end{align}

\begin{figure}
\centering
\includegraphics[scale=1]{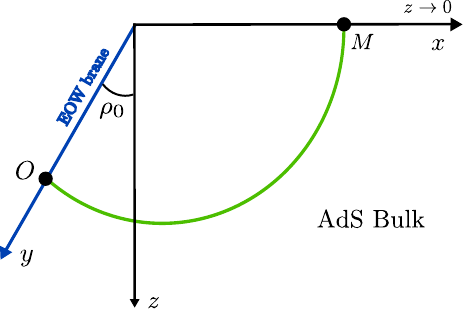}
\caption{Holographic dual of a BCFT$_2$ defined on a half plane $(x > 0)$. The EOW brane is considered to be at a constant hyperbolic angle $\rho_0$. (Figure borrowed from \cite{Dey:2025dlj}.)}\label{fig_bcft}
\end{figure}

The bulk AdS$_3$ geometry may be expressed in terms of AdS$_2$ foliations as follows
\begin{align}\label{met_fol}
ds^2 = d \rho^2 + \cosh ^2 \rho \left( \frac{d \tau^2 + dy^2}{y^2} \right),
\end{align}
where $y$ is the radial coordinate along the foliation and $\rho$ is the hyperbolic angle between the foliation and the normal to the asymptotic boundary. Note that we have assumed the AdS length scale to be unity for simplicity. \Cref{met_fol} may also be expressed in terms of the well-known Poincaré coordinates using
\begin{align}
x=-y \tanh \rho, \qquad z=y \sech \rho.
\end{align}
Assuming that the EOW brane is located at some constant $\rho = \rho_0$, the metric induced on this brane may be expressed as a conformally flat metric as
\begin{align}\label{met_brane}
ds^2 _{brane} = \Omega^2_y (d \tau^2 + dy^2) = \Omega^2_y \,\ ds^2 _{flat}, \qquad \Omega_y = \bigg| \frac{1}{y \sech \rho_0 } \bigg|.
\end{align}
where $\Omega_y$ is the conformal factor. From \cref{eq_kab,met_brane} we can then show that
\begin{align}
K_{ab} = \tanh \rho_0 \,\ h_{ab}, \qquad T = \tanh \rho_0.
\end{align}

The authors in \cite{Deng:2020ent,Chu:2021gdb} further developed the AdS$_3$/BCFT$_2$ setup by incorporating defect conformal matter localized on a tensionless EOW brane. This results in the brane relocating to a constant hyperbolic angle that depends on the induced tension. The bulk action in this modified defect AdS$_3$/BCFT$_2$ framework is then given by
\begin{align}
I = I_0 + \int_{Q} \sqrt{-h} ~ \mathcal{L}_Q,
\end{align}
where $\mathcal{L}_Q$ is the Lagrangian corresponding to the conformal matter on the brane.\footnote{See \cite{Chen:2020uac} for some examples.} Once again, by variation of this modified action with respect to the induced metric $h_{ab}$ and implementing the Neumann boundary condition, we arrive at \cite{Takayanagi:2011zk,Deng:2020ent,Chu:2021gdb}
\begin{align}
K_{ab} = (K-T)h_{ab}+8 \pi G_N \langle T_{ab} \rangle, \qquad \langle T_{ab} \rangle=\frac{2}{\sqrt{h}} \frac{\partial \mathcal{L} _Q}{\partial h_{ab}}.
\end{align}
A maximally symmetric geometry on the EOW brane leads to the expectation value of the stress-energy tensor of the conformal matter localized on the brane to satisfy $\langle T_{ab} \rangle \propto h_{ab}$. Consequently, its inclusion modifies only the tension of the brane, without backreacting on the bulk AdS$_3$ geometry. The EOW brane may then be regarded as a non-dynamical lower-dimensional defect embedded in the bulk spacetime.

\begin{figure}
\centering
\includegraphics[scale=0.78]{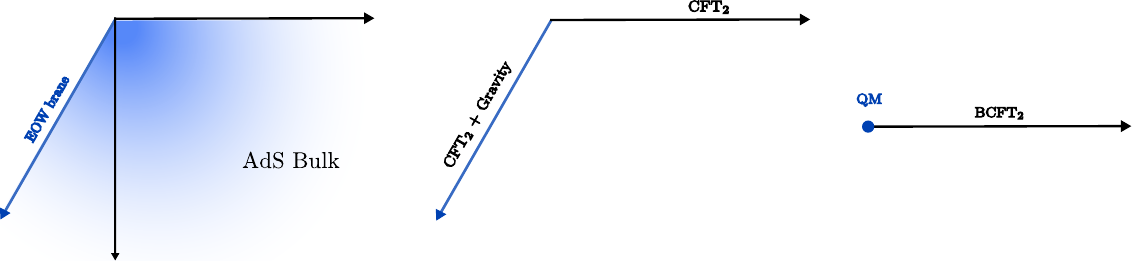}
\caption{This diagram represents a double holographic prescription. On the left, we have the bulk picture, in the center, we have the intermediate effective $2d$ picture, and on the right, we have the boundary picture.} 
\label{fig_dh}
\end{figure}

Additionally, through the double holographic prescription, an intermediate $2d$ description may be obtained by integrating out the bulk degrees of freedom, yielding an effective conformal field theory defined on the combined brane-boundary system. The sector localized on the brane is coupled to semiclassical gravity obtained via a partial dimensional reduction of the bulk geometry, while the remaining sector resides on a flat background. This intermediate description is itself holographically dual to a BCFT$_2$ coupled to a quantum mechanical system living on the boundary. The three complementary descriptions of double holography are illustrated in \cref{fig_dh}.

\subsection{Islands and Defect extremal surfaces}

In this section, we now review the island formula for the fine-grained entanglement entropy and the bulk DES prescription, which was introduced as the holographic counterpart of the island formula \cite{Deng:2020ent}. 

Consider a theory of semiclassical gravity coupled to a non-gravitating bath conformal field theory, where the gravitational sector may also contain some conformal matter or defects. In this framework, the fine-grained entropy associated with a single interval in the bath CFT may be computed using the island formula described in \cite{Almheiri:2019hni,Penington:2019npb, Almheiri:2019psf} as
\begin{align}\label{form_ee_is}
S_{Is}= \min_{X} \biggl[ \text{ext}_{X} \biggl\{ S_{eff}(A\cup I_A) +S_{area} (X) \biggr\} \biggr], \qquad X= \partial I_A
\end{align}
where $A$ refers to the subsystem on the radiation bath, $I_A$ is the corresponding island region on the brane, and $X$ denotes the boundary of the island $I_A$. Here $S_{eff}$ corresponds to the effective entanglement entropy of the subregion $A\cup I_A$, including contributions from conformal matter or defects localized on the brane.

From a holographic perspective, the DES formula was motivated by the observation that the gravity region relevant to determining the quantum extremal surface (QES) can be localized on the brane. This leads to the defect AdS$_3$/BCFT$_2$ framework described in \cref{ssec_adsbcft}, where certain geodesics with one endpoint on the defect, known as the defect extremal surface, may also be considered. In this case, the contribution from the bulk conformal matter located on the brane must be accounted for while computing the fine-grained entanglement entropy, which may then be described as
\begin{align}\label{form_ee_des}
S_{DES}= \min_{\Gamma, X} \biggl[ \text{ext}_{\Gamma,X} \biggl\{ S_{RT}(\Gamma)+S_{defect}(D) \biggr\} \biggr], \qquad X= \Gamma \cap D
\end{align}
where $\Gamma$ is a codimension-two RT surface homologous to $A$, and $D$ is the defect brane. $S_{defect}(D)$ then represents the entanglement entropy contribution arising from the conformal matter localized on the brane.


\subsection{AdS$_3$/ICFT$_2$ model}

We now review a class of $2d$ interface conformal field theories (ICFT$_2$s) introduced in \cite{Anous:2022wqh}, consisting of two CFT$_2$s defined on half-planes and coupled through a $1d$ interface. These boundary theories are denoted by CFT$^\text{I}$ and CFT$^\text{II}$, and possess central charges $\cI$ and $\cII$ respectively. The holographic dual is described by two locally AdS$_3$ spacetimes (specified by AdS$^\text{I}$ and AdS$^\text{II}$, with radii $L_\text{I}$ and $\LII$ respectively) which are separated by a permeable EOW brane.  

The bulk action describing this framework is given by \cite{Anous:2022wqh}
\begin{align}\label{ICFT-action}
I = & \frac{1}{16 \pi G_N} \left[ \int_{\mathcal{B}_\text{I}}\text{d}^3x \sqrt{-g_\text{I}} \left(R_\text{I} + \frac{2}{L_\text{I}^2} \right) + \int_{\mathcal{B}_\text{II}}\text{d}^3x \sqrt{-g_\text{II}} \left(R_\text{II} + \frac{2}{\LII^2} \right) \right] \notag \\
&~ + \frac{1}{8 \pi G_N} \left[ \int_\Sigma \text{d}^2 y \sqrt{-h} ( K_\text{I} - K_\text{II} )  - T \int_\Sigma \text{d}^2 y \sqrt{-h} \right] ,
\end{align}
where $h_{ab}$ is the induced metric on the EOW brane $\Sigma$ with tension $T$, while $\mathcal{B}_k$  ($k=\rm I, II$) represent the bulk AdS$^k$ geometry. The relative minus sign between extrinsic curvatures $K_\text{I}$ and $K_\text{II}$ on the EOW brane arises from choosing the outward normal to point from AdS$^\text{I}$ to AdS$^\text{II}$. Unlike the AdS/BCFT framework described in \cref{ssec_adsbcft}, where only the Neumann boundary conditions were applicable on the EOW brane, in the AdS/ICFT setup, two junction conditions apply: continuity of the induced metric across the brane, and the Israel junction condition given as
\begin{align} \label{Israel-ICFT}
\big( K_{\text{I},ab} - K_{\text{II},ab} \big) - h_{ab} \big( K_{\text{I}} - K_{\text{II}} \big) = -T ~ h_{ab}.
\end{align}

The bulk AdS$_3$ geometries may once again be expressed as AdS$_2$ foliations given by 
\begin{align}\label{metric-ICFT-AdS2-slicing}
\text{d}s_{\mathcal{B}_{k}}^2 &= d\rho_{k}^2 + L_{k}^2 \cosh^2 \left( \frac{\rho_{k}}{L_{k}} \right) \frac{-dt_{k}^2 + dy_{k}^2}{y_{k}^2}.
\end{align}
Considering that the EOW brane is located at $\rho_k=\rho_k^0$, the induced metric is given as 
\begin{align}\label{eq_cfactor}
\text{d}s_{\Sigma}^2  = \Omega_k(y)^2 (-\text{d} t_{k}^2 + \text{d} y_{k}^2), \qquad \Omega_k(y) = \left|  \frac{ L_{k} \cosh \left( \frac{\rho_{k}^0}{L_{k}} \right)}{y_k}  \right|,
\end{align}
which is simply a $2d$ flat geometry with conformal factor $\Omega_k(y)$. Continuity of the induced metric across the two patches identifies the bulk coordinates $(t_{\rm I},y_{\rm I})$ in AdS$^{\rm I}$ and $(t_{\rm II},y_{\rm II})$ in AdS$^{\rm II}$, with the constraint
\begin{align}\label{JC2}
\LI^2 \cosh^2 \left( \frac{\rhoI}{\LI} \right) = \LII^2 \cosh^2 \left( \frac{\rhoII}{\LII} \right),
\end{align}
while the Israel junction condition in \cref{Israel-ICFT} gives 
\begin{align}\label{eq_tension}
T = \frac{1}{\LI} \tanh \left( \frac{\rhoI}{\LI} \right) + \frac{1}{\LII} \tanh \left( \frac{\rhoII}{\LII} \right).
\end{align}
Solving the two conditions \cref{JC2,eq_tension} fixes the location of the brane in the two AdS$_3$ geometries as \cite{Anous:2022wqh}
\begin{align}\label{Branelocation}
\tanh \left( \frac{\rhoI}{\LI} \right) = \frac{\LI}{2 T} \left( T^2 + \frac{1}{\LI^2} - \frac{1}{\LII^2} \right), \qquad \tanh \left( \frac{\rhoII}{\LII} \right) = \frac{\LII}{2 T} \left( T^2 - \frac{1}{\LI^2} + \frac{1}{\LII^2} \right).
\end{align}
\Cref{eq_tension} also provides the lower and upper limit to the brane tension $T$ as
\begin{align}
\bigg| \frac{1}{\LI} - \frac{1}{\LII} \bigg| < T < \frac{1}{\LI} + \frac{1}{\LII}.
\end{align}
The large tension limit corresponds to the case where the brane approaches the asymptotic boundary of both AdS$_3$ geometries. An intermediate effective $2d$ description of the AdS$_3$/ICFT$_2$ framework may be described in this regime, which shall be discussed in detail in \cref{sec_desinterface}.


\section{Defect extremal surfaces and interface CFTs}\label{sec_desinterface}

Having discussed the DES prescription and the AdS$_3$/ICFT$_2$ framework, we now incorporate non-dynamical conformal defects on the EOW brane (along the lines of \cite{Deng:2020ent}), which allow us to describe an effective lower-dimensional CFT$_2$s on the brane-boundary combination, partially coupled to semiclassical gravity. We derive the effective lower-dimensional gravity on the EOW brane through partial dimensional reduction of the bulk geometry, and subsequently generalize the defect extremal surface formula for the fine-grained entanglement entropy to interface CFTs.


\subsection{Defect AdS$_3$/ICFT$_2$ model and holography}

In this section, we shall describe the defect AdS$_3$/ICFT$_2$ framework, which arises on incorporating bulk conformal matter on the EOW brane. The bulk action in \cref{ICFT-action} is then modified as
\begin{align}
I \to I + \int_{\Sigma} \sqrt{-h} ~ \mathcal{L}_\Sigma,
\end{align}
where $\mathcal{L}_\Sigma$ is the Lagrangian corresponding to the conformal matter on the brane $\Sigma$. Once again, two different junction conditions apply on the brane -- the continuity of the induced metric across the brane and the Israel junction condition given as
\begin{align} \label{Israel-defectICFT}
\big( K_{\text{I},ab} - K_{\text{II},ab} \big) - h_{ab} \big( K_{\text{I}} - K_{\text{II}} - T\big) = 8 \pi G_N \langle T_{ab} \rangle .
\end{align}
Maximal symmetry on the EOW brane once again leads to $\langle T_{ab} \rangle \propto h_{ab}$. The conformal matter localized on the brane then simply modifies the brane tension $T$ without backreacting on the bulk AdS$_3$ geometries, and the EOW brane may once again be regarded as a non-dynamical lower-dimensional defect embedded in the bulk spacetime.

An intermediate lower dimension theory may be obtained on the brane-boundary combination via partial reduction of the bulk degrees of freedom in the large tension limit, where the brane approaches the asymptotic boundaries of both the AdS$_3$ geometries. The dynamics of the defect matter on the brane become locally conformal, resulting in a weakly gravitating system localized on the brane, coupled to the original CFT$_2$s. On the asymptotic boundary, the CFTs still reside on a flat background. This is illustrated in \cref{fig_icft}. In this limit, the dual to the semiclassical $2d$ island formula for the generalized entanglement entropy is given by the Ryu-Takayanagi prescription.


\subsection{Effective theory on brane-boundary combination}

\begin{figure}
\centering
\includegraphics[scale=0.78]{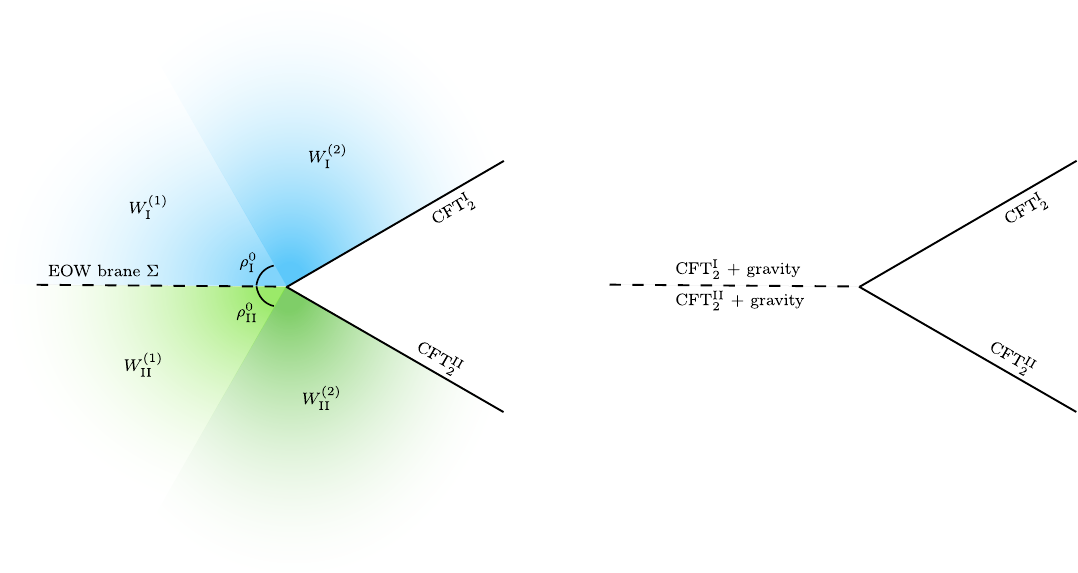}
\caption{On the left, we have the schematics for partial reduction of the AdS$_3$/ICFT$_2$ setup with the EOW brane $\Sigma$. The diagram on the right represents the lower-dimensional effective $2d$ description, where some of the CFTs are weakly coupled to gravity.} 
\label{fig_icft}
\end{figure}

We now discuss the emergence of a lower-dimensional gravity on the EOW brane via a partial dimensional reduction of the bulk AdS$_3$ geometry, and describe the effective lower-dimensional theory described on the brane-boundary combination.\footnote{Note that the following analysis may be reproduced from those in \cite{Afrasiar:2023nir} by setting the brane fluctuations to zero.}

As depicted in \cref{fig_icft}, the EOW brane is fixed at 
\begin{align}
\rho_{\text{I}}=\rhoI, \qquad \rho_{\text{II}}=\rhoII,
\end{align}
with respect to the normals to the asymptotic boundary. Note that we consider large $\rho_k^0$ corresponding to the large tension limit. The two bulk AdS$_3$ geometries on either side of this EOW brane may be divided into two wedges $W_k^{(1)}$ and $W_k^{(2)}$ such that
\begin{align}
\mathcal{B}_k = W_k^{(1)} \cup W_k^{(2)}.
\end{align}
From the AdS$_3$/CFT$_2$ correspondence, the wedges $W_k^{(2)}$ are dual to non-gravitating CFT$_2$s on the two asymptotic boundaries. On the other hand, integrating out the bulk degrees of freedom on the wedges $W_k^{(1)}$ along the $\rho_k$ direction allows us to obtain the lower-dimensional gravity on the EOW brane. The dimensional reduction of the bulk Einstein-Hilbert terms in \cref{ICFT-action} then simplifies to
\begin{align} \label{EH-reduction-nofluct}
I_{\rm EH} &= \frac{1}{16\pi G_N} \int_{W_{\rm I}^{(1)}} d^3x \sqrt{-g_{\rm I}} \left[ R_{\rm I} + \frac{2}{\LI^2} \right] + \frac{1}{16\pi G_N} \int_{W_{\rm II}^{(1)}} d^3x \sqrt{-g_{\rm II}} \left[ R_{\rm II} + \frac{2}{\LII^2} \right] \notag\\
& = \frac{1}{16\pi G_N} \int_{\Sigma} d^2y \sqrt{-h} \Bigg[ \left(\rhoI+\rhoII\right)R^{(2)} 	-\frac{2}{\LI} \tanh\!\left(\frac{\rhoI}{\LI}\right) -\frac{2}{\LII} \tanh\!\left(\frac{\rhoII}{\LII}\right) \Bigg],
\end{align}
where $h_{ab}$ is the metric induced on the EOW brane $\Sigma$. Using the extrinsic curvatures on the brane
\begin{align}
K_{{\rm I},ab} = \frac{1}{\LI} \tanh\!\left(\frac{\rhoI}{\LI}\right)h_{ab}, \qquad K_{{\rm II},ab} = -\frac{1}{\LII} \tanh\!\left(\frac{\rhoII}{\LII}\right)h_{ab},
\end{align}
the Gibbons-Hawking and brane tension terms in \cref{ICFT-action} then become
\begin{align} \label{GH-reduction-nofluct}
I_{\rm GH+T} &= \frac{1}{8\pi G_N} \int_\Sigma d^2y \sqrt{-h} \left[ K_{\rm I}-K_{\rm II}-T \right] \notag\\
&= \frac{1}{8\pi G_N} \int_\Sigma d^2y \sqrt{-h} \left[ \frac{1}{\LI} \tanh\!\left(\frac{\rhoI}{\LI}\right) + \frac{1}{\LII} \tanh\!\left(\frac{\rhoII}{\LII}\right) \right],
\end{align}
where we have utilized \cref{eq_tension}. Adding \eqref{EH-reduction-nofluct} and \eqref{GH-reduction-nofluct} yields
\begin{align}
I_{\rm eff} = \frac{\rhoI+\rhoII}{16\pi G_N} \int_\Sigma d^2y \sqrt{-h}\,R^{(2)} = \frac{1}{16\pi G_N^{(2)}} \int_\Sigma d^2y \sqrt{-h}\,R^{(2)} .
\end{align}
where we have defined the effective two-dimensional Newton constant on the EOW brane as
\begin{align}\label{eq_g2}
\frac{1}{G_N^{(2)}} = \frac{\rhoI+\rhoII}{G_N}.
\end{align}


\subsection{Generalized DES formula for entanglement entropy}

We now describe the island prescription and the generalized DES formula for the fine-grained entanglement entropy in the defect AdS$_3$/ICFT$_2$ framework discussed above. Consider a QFT coupled to gravitational theory on a hybrid manifold $\mathcal{M} = \Sigma \cup \mathcal{M}_{\rm I} \cup \mathcal{M}_{\rm II}$ such that $\partial \mathcal{B}_k \equiv \Sigma \cup \mathcal{M}_k$, where the EOW brane $\Sigma$ joins smoothly to the two non-gravitational manifolds $\mathcal{M}_k$ on the asymptotic boundary. We consider a single subsystem $A$ on this hybrid manifold with density matrix $\rho_A$.

Unlike earlier works involving defect extremal surfaces where the EOW brane was coupled to a single radiation bath \cite{Deng:2020ent}, the presence of two baths modify the structure of the dominant replica wormhole saddle and provide two independent mechanisms for the origin of the island region in the semiclassical description.\footnote{The two scenarios are discussed in greater details in \cite{Afrasiar:2023nir}.}

For a subsystem $A = A^{\rm I} \cup A^{\rm II}$ with $A^k \subset \mathcal{M}_k$, the corresponding island region $I_A$ on the EOW brane depends on the degrees of freedom of both the CFT baths. Consequently, we have $I_A^{\rm I} = I_A^{\rm II} = I_A$, where the island region $I_A^k$ depends on the degrees of freedom of CFT$_2^k$. The density matrix in the effective theory then factorizes as
\begin{align}
\rho_{A \cup I_A} \sim \rho_{A^{\rm I} \cup I_A} \otimes \rho_{A^{\rm II} \cup I_A}.
\end{align}
From the double holographic perspective, this saddle corresponds to a scenario in which the bulk RT surfaces homologous to subsystem $A$ are composed of geodesics that cross from AdS$^{\rm I}$ to AdS$^{\rm II}$. This is depicted in \cref{fig_tii} and is also known as the island phase.

On the other hand, if the subsystem $A$ resides entirely in the bath $\mathcal{M}^{\rm I}$, with central charges $\cI > \cII$, then depending on the size and location of $A$ we may observe a corresponding island $I_A$ on the EOW brane $\Sigma$. Given that $\Sigma$ is common to both the CFTs in the intermediate lower-dimensional picture, the CFT$^{\rm II}$ degrees of freedom also conceive an induced island $I_A^{\rm (II/I)}$. In this case, the density matrix is factorized as
\begin{align}
\rho_{A \cup I_A} \sim \rho_{A^{\rm I} \cup I_A} \otimes \rho_{I_A^{\rm (II/I)}}.
\end{align}
From the double holographic perspective, this corresponds to a scenario where the RT surface homologous to $A$ double-crosses the EOW brane, such that the minimal geodesic originating in AdS$^{\rm I}$ penetrates into AdS$^{\rm II}$ and then returns back. This case is also known as the induced island phase and is illustrated in \cref{fig_tiii}.

Given the scenarios discussed above, we are now in a position to describe the island formula for the fine-grained entanglement entropy for the subsystem $A$ in the defect AdS$_3$/ICFT$_2$ framework as
\begin{align}\label{eq_icft_is}
S_{Is}= \min_{X} \biggl[ \text{ext}_{X} \biggl\{ S_{eff}(A\cup I_A) +S_{area} (X) \biggr\} \biggr], \qquad X= \partial I_A
\end{align}
where the first term in the curly brackets is the effective entanglement entropy of $A$ and its corresponding island $I_A$, and is given as
\begin{align}
S_{eff}(A \cup I_A) = 
\begin{cases}
S_{eff}(A^{\rm I} \cup I_A) + S_{eff}(A^{\rm II} \cup I_A), & \qquad \text{island phase} \\
S_{eff}(A^{\rm I} \cup I_A) + S_{eff}(I_A^{\rm (II/I)}), & \qquad \text{induced island phase}
\end{cases}
\end{align}
while the second term is the area term given by
\begin{align}\label{def_s_area}
S_{area} (X) = \frac{Area(\partial I_A)}{4 G_N^{(2)}}.
\end{align}

From the bulk perspective, the entanglement entropy may be determined using the generalized DES formula in the defect AdS$_3$/ICFT$_2$ framework, given as
\begin{align}\label{eq_icft_des}
S_{DES}= \min_{\Gamma, X} \biggl[ \text{ext}_{\Gamma,X} \biggl\{ S_{RT}(\Gamma)+S_{defect}(D) \biggr\} \biggr], \qquad X= \Gamma \cap D.
\end{align}
Here $\Gamma = \Gamma^{\rm I} \cup \Gamma^{\rm II}$, where $\Gamma^k$ correspond to the bulk geodesic(s) in AdS$^k$. $S_{defect}$ is once again the contribution to the entanglement entropy due to bulk conformal matter localized on the EOW brane.

\subsection*{The area and defect terms}

If we consider the entanglement entropy island on the EOW brane to be described by the subsystem $I_A \equiv [y_1, y_2]$ on a constant time slice, the area and defect terms in \cref{eq_icft_is,eq_icft_des} respectively is given as  
\begin{align}\label{eq_s_area}
S_{area} = \frac{\cI}{3} \rhoI + \frac{\cII}{3} \rhoII
\end{align}
and
\begin{align}\label{eq_s_defect}
S_{defect} = 
\begin{cases}
  \frac{\cI}{6} \log \frac{\LI^2 (y_2-y_1)^2}{y_1 y_2 \epsilon_y^2 \sech ^2 \rhoI} + \frac{\cII}{6} \log \frac{\LII^2 (y_2-y_1)^2}{y_1 y_2 \epsilon_y^2 \sech ^2 \rhoII} \qquad & \eta <1, \\
  \frac{\cI}{3} \log \frac{2 \LI}{\epsilon_y \sech \rhoI} + \frac{\cII}{3} \log \frac{2 \LII}{\epsilon_y \sech \rhoII} \qquad & \eta >1
\end{cases}
\end{align}
where $\eta = \frac{(y_2-y_1)^2}{4 y_2 y_1}$ is the cross ratio. The regime described by $\eta < 1$ corresponds to the channel where the operator product expansion (OPE) dominates over the boundary operator expansion (BOE), and vice versa for $\eta > 1$. $S_{defect}$ transitions from one dominant channel to the other at the critical value $\eta_c = 1$. The detailed computations are provided in \cref{app_sad}.

Additionally, we have rescaled $\rho_k^0 / L_k \to \rho_k^0$ for simplicity, a convention which shall be adopted for the rest of the article.


\section{Time independent scenario}\label{sec_TI}

\begin{figure}[t]
	\centering
	\begin{subfigure}{0.45\linewidth}
		\includegraphics[scale=0.8]{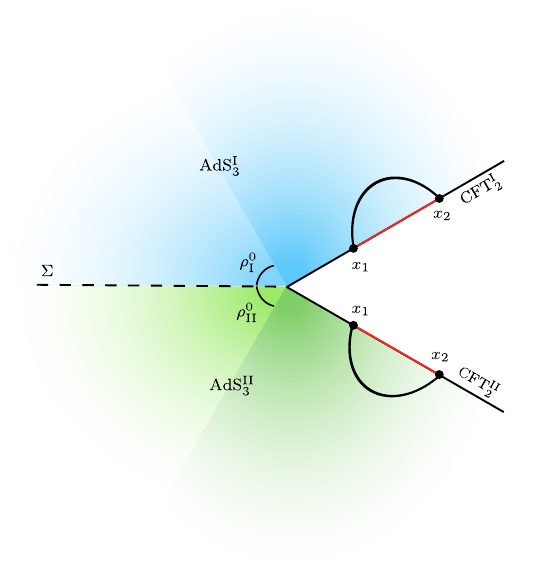}\caption{No-Island Phase}
		\label{fig_tini}
	\end{subfigure}
	\hfill
	\begin{subfigure}{0.45\linewidth}
		\includegraphics[scale=0.8]{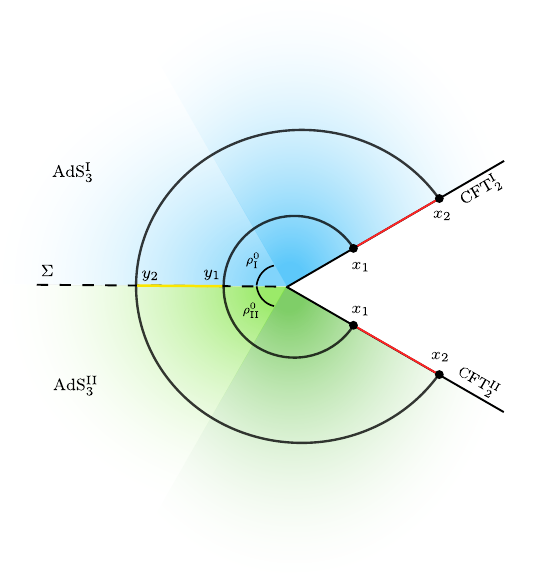}\caption{Island Phase}
		\label{fig_tii}
	\end{subfigure}
	\begin{subfigure}{0.45\linewidth}
		\includegraphics[scale=0.8]{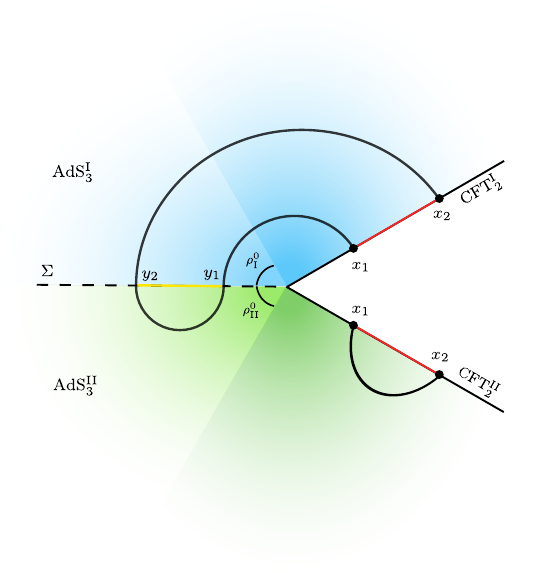}\caption{Induced Island Phase}
		\label{fig_tiii}
	\end{subfigure}
	\caption{The various entanglement entropy phases in the time-independent scenario. The subsystem $A$ (represented in red) is considered on both the CFTs. The corresponding island regions in the island and induced island phases are located on the EOW brane and are depicted in yellow. The black curves represent the homologous bulk RT surfaces.}
	\label{fig_ti}
\end{figure}

In this section, we consider subsystem $A \equiv [x_1,x_2]_{\rm I} \cup [x_1,x_2]_{\rm II}$ described on both the CFT$_2$s on a constant time slice and compute the entanglement entropy. From the field theoretic perspective, the entanglement entropy is computed using the island formula in \cref{eq_icft_is}, while the holographic entanglement entropy is calculated using the generalized bulk DES formula given in \cref{eq_icft_des}. We observe three distinct phases --- the no-island phase, the island phase, and the induced island phase, as illustrated in \cref{fig_ti}. We shall now discuss them in detail.

\subsection{No Island Phase}

\paragraph{Field Theory Computations} This phase corresponds to the scenario where the subsystems are small and far away from the interface, as illustrated in \cref{fig_tini}. The entanglement entropy is then computed using the island formula in \cref{eq_icft_is}, where the area term $S_{area}$ vanishes due to the absence of any entanglement entropy island on the brane. The entanglement entropy for the CFT$_2^{\rm I}$ may then be expressed as
\begin{align}
S^{\rm (I)} = \lim_{n \to 1} \frac{1}{1-n}  \log \langle \sigma(x_1) \bar{\sigma}(x_2) \rangle_{_{\text{CFT}_2^{\rm I}}}  =  \frac{\cI}{3} \log \frac{x_2-x_1}{\epsilon},
\end{align}
where in the second step we have utilized the standard form of the CFT$_2$ two-point correlator
\begin{align}\label{eq_2pf}
\langle \sigma(z_1) \bar{\sigma}(z_2) \rangle = \frac{1}{|z_2-z_1|^{2 h_n}}.
\end{align}
Similarly for CFT$_2^{\rm II}$ we may obtain 
\begin{align}
S^{\rm (II)} = \frac{\cII}{3} \log \frac{x_2-x_1}{\epsilon},
\end{align}
and the generalized entanglement entropy, which is also the final entanglement entropy in this phase, is given as
\begin{align}\label{eq_ti_ni_cft}
S = S_{gen} = \frac{\cI+\cII}{3} \log \frac{x_2-x_1}{\epsilon}.
\end{align}

\paragraph{Bulk Computations} From the bulk perspective, the RT surfaces homologous to the subsystem A are depicted by the black curves in \cref{fig_tini}. Since the RT surfaces do not intersect the EOW brane, the defect term $S_{defect}$ does not contribute to the generalized entanglement entropy in this phase. The final holographic entanglement entropy is then obtained using \cref{,eq_rt,eq_geol,eq_ip,eq_icft_des} in the large tension limit as
\begin{align}
S = \frac{\LI+ \LII}{2 G_N} \log \frac{x_2-x_1}{\epsilon}, 
\end{align}
which matches with the field theory results in \cref{eq_ti_ni_cft} on applying the Brown-Henneaux relation $c_k = \frac{3 L_k}{2 G_N}$ \cite{Brown:1986nw}.


\subsection{Island Phase}\label{ssec_tii}

\paragraph{Field Theory Computations} This phase corresponds to the scenario where entanglement entropy islands are observed on the EOW brane and are described as $[y_1,y_2]_{\rm I,II}$ on a constant time slice, as depicted in \cref{fig_tii}. The effective entanglement entropy term in \cref{eq_icft_is} may be expressed in terms of the following four-point twist field correlators on CFT$^{\rm I}_2$, which in the large central charge limit factorize as
\begin{align}\label{eq_tii_factorize}
\langle \sigma(x_1) \bar{\sigma}(x_2) \sigma(y_2) \bar{\sigma}(y_1) \rangle_{_{\text{CFT}_2^{\rm I}}} \sim \langle \sigma(x_1) \bar{\sigma}(y_1) \rangle_{_{\text{CFT}_2^{\rm I}}} \langle \bar{\sigma}(x_2) \sigma(y_2)  \rangle_{_{\text{CFT}_2^{\rm I}}}.
\end{align}
The effective terms on CFT$^{\rm I}$ is then given as
\begin{align}
S^{\rm (I)} & = \lim_{n \to 1} \frac{1}{1-n} \left[ \log \Omega_{\rm I}(y_1)^{-2 h_n^{\rm I}} \langle \sigma(x_1) \bar{\sigma}(y_1) \rangle_{_{\text{CFT}_2^{\rm I}}} + \log \Omega_{\rm I}(y_2)^{-2 h_n^{\rm I}} \langle \bar{\sigma}(x_2) \sigma(y_2) \rangle_{_{\text{CFT}_2^{\rm I}}} \right] \notag \\
& = \frac{\cI}{6} \left(\log \frac{(x_1+y_1)^2}{\epsilon \epsilon_y} + \log \frac{(x_2+y_2)^2}{\epsilon \epsilon_y} + \log \frac{\LI}{ y_1 \sech \rhoI} + \log \frac{\LI}{ y_2 \sech \rhoI} \right) ,
\end{align}
such that $h_n^k = \frac{c_k}{24}\left( n-1/n \right)$ is the conformal weights in CFT$_2^k$, and the conformal factors $\Omega_{\rm I}(y_1),\Omega_{\rm I}(y_2)$ are given in \cref{eq_cfactor}. The four-point correlator in CFT$^{\rm II}$ factorizes in a fashion similar to \cref{eq_tii_factorize}, and the entanglement entropy $S^{\rm (II)}$ is obtained by substituting $(\cI,\LI,\rhoI) \to (\cII,\LII,\rhoII)$ in the above expression. Together with the area term in \cref{eq_s_area}, we obtain the generalized entanglement entropy, which is extremized with respect to $y_1,y_2$ at
\begin{align}
y_1 = x_1, \qquad y_2 = x_2.
\end{align}
Substituting them back into $S_{gen}$, the final entanglement entropy is obtained as
\begin{align}\label{eq_ti_i_cft}
S & = \frac{\cI}{6} \left( 2 \rhoI + \log \frac{2 x_1}{\epsilon} + \log \frac{2 x_2}{\epsilon} +2 \log \frac{2 \LI}{\epsilon_y \sech \rhoI} \right) \notag \\
& \qquad + \frac{\cII}{6} \left( 2 \rhoII + \log \frac{2 x_1}{\epsilon} + \log \frac{2 x_2}{\epsilon} +2 \log \frac{2 \LII}{\epsilon_y \sech \rhoII} \right).
\end{align}

\paragraph{Bulk Computations} From the bulk perspective, the RT surfaces homologous to the subsystem $A$ are illustrated by the black curves in \cref{fig_tii}, which cross the interface from AdS$^{\rm I}$ to AdS$^{\rm II}$. Using \cref{eq_rt,eq_geol,eq_ip} and the DES formula in \cref{eq_icft_des}, the generalized entanglement entropy may be obtained as
\begin{align}
S_{gen} & = \frac{\LI}{4G_N} \left( \log \frac{(x_1^2 + y_1^2) + 2 x_1 y_1 \tanh \rhoI}{\epsilon y_1 \sech \rhoI} + \log \frac{(x_2^2 + y_2^2) + 2 x_2 y_2 \tanh \rhoI}{\epsilon y_2 \sech \rhoI} \right) \\
& \qquad + \frac{\LII}{4G_N} \left( \log \frac{(x_1^2 + y_1^2) + 2 x_1 y_1 \tanh \rhoII}{\epsilon y_1 \sech \rhoII} + \log \frac{(x_2^2 + y_2^2) + 2 x_2 y_2 \tanh \rhoII}{\epsilon y_2 \sech \rhoII} \right) + S_{defect}\notag 
\end{align}
where $S_{defect}$ is given by \cref{eq_s_defect}. 

Recall that $S_{defect}$ has two different expressions depending on the value of the cross ratio $\eta$. Substituting the expression corresponding to $\eta < 1$, we observe that $\partial_{y_1}S_{gen},\partial_{y_2}S_{gen} <0$ for all values of $y_1,y_2$, implying the absence of any extremal solution. On the other hand, $\eta > 1$ leads to an extremal solution at $(y_1,y_2) = (x_1,x_2)$. Substituting them back into $S_{gen}$, we finally obtain the holographic entanglement entropy as \cref{eq_ti_i_cft} upon implementing the Brown-Henneaux relation and the large tension limit.


\subsection{Induced Island Phase}\label{ssec_tiii}

\paragraph{Field Theory Computations} In this phase an entanaglement entropy island corresponding to subsystem $A$ is observed in CFT$^{\rm I}$, while an induced island is observed in CFT$^{\rm II}$. As given in \cref{fig_tiii}, the location of the island region on the EOW brane is once again considered to be $[y_1,y_2]_{\rm I,II}$ on a constant time slice. The effective term in \cref{eq_icft_is} may be computed in terms of four-point twist field correlators on CFT$^{\rm I,II}$, which in the large central charge limit factorize as 
\begin{align}
\langle \sigma(x_1) \bar{\sigma}(x_2) \sigma(y_2) \bar{\sigma}(y_1) \rangle_{_{\text{CFT}_2^{\rm I}}} & \sim \langle \sigma(x_1) \bar{\sigma}(y_1) \rangle_{_{\text{CFT}_2^{\rm I}}} \langle \bar{\sigma}(x_2) \sigma(y_2)  \rangle_{_{\text{CFT}_2^{\rm I}}}, \notag \\
\langle \sigma(x_1) \bar{\sigma}(x_2) \sigma(y_2) \bar{\sigma}(y_1) \rangle_{_{\text{CFT}_2^{\rm II}}} & \sim \langle \sigma(x_1) \bar{\sigma}(x_2) \rangle_{_{\text{CFT}_2^{\rm II}}} \langle \bar{\sigma}(y_1) \sigma(y_2)  \rangle_{_{\text{CFT}_2^{\rm II}}}.
\end{align}
The effective entanglement entropy for CFT$^{\rm I}$ is then given as 
\begin{align}\label{eq_tiii_s1}
S^{\rm (I)} & = \lim_{n \to 1} \frac{1}{1-n} \left[ \log \Omega_{\rm I}(y_1)^{-2 h_n^{\rm I}} \langle \sigma(x_1) \bar{\sigma}(y_1) \rangle_{_{\text{CFT}_2^{\rm I}}} + \log \Omega_{\rm I}(y_2)^{-2 h_n^{\rm I}} \langle \bar{\sigma}(x_2) \sigma(y_2) \rangle_{_{\text{CFT}_2^{\rm I}}} \right] \notag \\
& = \frac{\cI}{6} \left(\log \frac{(x_1+y_1)^2}{\epsilon \epsilon_y} + \log \frac{(x_2+y_2)^2}{\epsilon \epsilon_y} + \log \frac{\LI}{ y_1 \sech \rhoI} + \log \frac{\LI}{ y_2 \sech \rhoI} \right),
\end{align}
while for CFT$^{\rm II}$ we have 
\begin{align}\label{eq_tiii_s2}
S^{\rm (II)} & =  \lim_{n \to 1} \frac{1}{1-n} \left[ \log \Omega_{\rm II}(y_1)^{-2 h_n^{\rm II}} \langle \sigma(x_1) \bar{\sigma}(x_2) \rangle_{_{\text{CFT}_2^{\rm II}}} + \log \Omega_{\rm II}(y_2)^{-2 h_n^{\rm II}} \langle \bar{\sigma}(y_1) \sigma(y_2) \rangle_{_{\text{CFT}_2^{\rm II}}} \right] \notag \\
& = \frac{\cII}{6} \left( 2 \log \frac{x_2-x_1}{\epsilon} + 2 \log \frac{y_2-y_1}{\epsilon_y} + \log \frac{\LII}{ y_1 \sech \rhoII} + \log \frac{\LII}{ y_2 \sech \rhoII} \right).
\end{align}
The generalized entanglement entropy $S_{gen}$ is then obtained as the sum of the above expressions and the area term in \cref{eq_s_area}. On extremization of $S_{gen}$ with respect to $y_1,y_2$ we obtain
\begin{align}
(y_1,y_2) \to (k_s x_1, \frac{x_2}{k_s}), \qquad \text{such that } k_s = \frac{\cI + \cII}{\cI - \cII},
\end{align}
for $x_2>>x_1$ and $\cI >> \cII$ (see \cref{app_kcurve}). We finally obtain the entanglement entropy as 
\begin{align}\label{eq_ti_ii_cft}
S & = \frac{\cI}{6} \left( 2 \rhoI + \log \frac{x_1}{\epsilon} \frac{(1+k_s)^2}{2 k_s} + \log \frac{x_2}{\epsilon} \frac{(1+k_s)^2}{2 k_s} + 2 \log \frac{2 \LI}{\epsilon_y \sech \rhoI} \right) \notag \\
& \qquad + \frac{\cII}{6} \left( 2 \rhoII + 2 \log \frac{x_2 - x_1}{\epsilon} + 2 \log \frac{\frac{x_2}{k_s}-k_s x_1}{2 \sqrt{x_2 x_1}} + 2 \log \frac{2 \LII}{\epsilon_y \sech \rhoII} \right).
\end{align}

\paragraph{Bulk Computations} From the bulk perspective, the RT surfaces homologous to subsystem $A$ are given by the black curves in \cref{fig_tiii}. The RT surface in this phase double crosses the EOW brane from AdS$^{\rm I}$ to AdS$^{\rm II}$ and back, which leads to an island region on AdS$^{\rm I}$ and an induced island region on AdS$^{\rm II}$. The  holographic entanglement entropy may be computed using \cref{eq_icft_des}, where the effective term is given by the sum of the following
\begin{align}
S^{\rm (I)} & = \frac{\LI}{4G_N} \left( \log \frac{(x_1^2 + y_1^2) + 2 x_1 y_1 \tanh \rhoI}{\epsilon y_1 \sech \rhoI} + \log \frac{(x_2^2 + y_2^2) + 2 x_2 y_2 \tanh \rhoI}{\epsilon y_2 \sech \rhoI} \right), \notag \\
S^{\rm (II)} & = \frac{\LII}{4G_N} \left( \cosh ^{-1} \frac{y_1^2 + y_2^2 + (y_2-y_1)^2 \sinh ^2 \rhoII}{2 y_2 y_1} + 2 \log \frac{x_2 - x_1}{\epsilon} \right),
\end{align}
and the generalized entanglement entropy is obtained by adding the defect term in \cref{eq_s_defect}. Once again we assume $x_2>>x_1, \cI >> \cII$ and the large tension limit. In this limit, we arrive at an unphysical solution for $\eta<1$, while $S_{gen}$ is extremized at $(y_1,y_2) \to (k_s x_1, \frac{x_2}{k_s})$ with $k_s = \frac{\cI + \cII}{\cI - \cII}$ for $\eta>1$. The holographic entanglement entropy may be finally obtained as \cref{eq_ti_ii_cft} on applying the Brown-Henneaux relation in the large tension limit.


\section{Time dependent scenario}\label{sec_TD}

In this section, we review how a $2d$ eternal black hole emerges from a time-dependent defect AdS$_3$/ICFT$_2$ model, along the lines of \cite{Fujita:2011fp,Li:2021dmf,Chu:2021gdb,Rozali:2019day}. As discussed earlier, the interface separating the CFT$_2$s on the asymptotic boundary is dual to an EOW brane in the bulk, separating the two AdS$_3$ dual bulk geometries. In the time-dependent case, the interface is described at $\tau = 0$ with the CFTs described on half planes given by $(x,\tau>0)$. The EOW brane is located at constant hyperbolic angles $\rho_k^0$ and described by the equation $\tau=-z \sinh \rho_k^0$ in the bulk AdS$^k$ geometry, which are given as
\begin{align}
ds^2 & = L_k ^2 \left( d\rho_k^2+ \cosh ^2 \rho_k \left( \frac{dx^2+dy^2}{y^2} \right) \right), \notag \\
& = \frac{L_k ^2}{z^2} (d \tau ^2+dx^2+dz^2).
\end{align}
Using the set of global conformal transformations
\begin{align}\label{eq_trans1}
& \tau = \frac{2 L_k (\tau'^2 + x'^2 + z'^2 - L_k^2)}{(\tau' + L_k)^2 + x'^2 + z'^2}, \qquad x = \frac{4 L_k^2 x'}{(\tau' + L_k)^2 + x'^2 + z'^2}, \notag \\ 
& \hspace{3cm} z = \frac{4 L_k^2 z'}{(\tau' + L_k)^2 + x'^2 + z'^2},
\end{align}
the interface is mapped to the circles ${x'}^2+{\tau '}^2=L_k^2$, while the EOW brane is now a part of spheres described by $(z'+L_k \sinh \rho_k^0)^2+{x'}^2+{\tau '}^2=L_k^2 \cosh ^2 \rho_k^0$.  A Lorentzian solution may be obtained by analytically continuing $\tau' \to it'$, upon which the geometry of the EOW brane becomes part of a hyperboloid.

The left and right boundaries of the EOW brane in the bulk AdS$^k$ geometry are described by $ x' = \pm \sqrt{ t'^2 +L_k^2} $ in the Lorentzian solution. The light curves at $ t' \to \pm \infty $ then correspond to horizons of $2d$ eternal black holes on the EOW brane, which are given by \begin{align}
x' = \pm t', \qquad \qquad z' = L_k e^{-\rho_k^0 },
\end{align}
with the black hole interior being specified by $ |x'| < t' $. Though the two regions on the AdS$_3$ asymptotic boundary are causally disconnected, they are still connected in the bulk AdS$_3$ geometry. This indicates quantum entanglement between the two regions on the asymptotic boundary, as also described in \cite{Maldacena:2001kr} for the case of general eternal AdS-Schwarzschild black holes.

The two-sided eternal black hole then emerges on the EOW brane as part of the effective $2d$ description via a partial Randall-Sundrum reduction of the bulk degrees of freedom, and is coupled to the bath CFT$^{\rm I,II}$s. To describe the time evolution, it is possible to capture the near-horizon geometry of the black hole by introducing the Rindler coordinates $(X,T)$ as\footnote{Note that the Rindler coordinates correspond to an observer moving with a constant acceleration in the Lorentzian geometry described above. A constant-accelerated observer in Rindler space corresponds to a static observer in Schwarzschild spacetime.}
\begin{align}\label{eq_trans21}
x' = L_k e^X \cosh T, \qquad \tau' = i L_k e^X \sinh T
\end{align}
This transformation may be applied to both the left and right patches, ultimately resulting in a two-sided $2d$ eternal black hole coupled to the bath CFT.

In the course of the transformations in \cref{eq_trans1,eq_trans21}, the UV cut-off transforms as
\begin{align}\label{eq_trans22}
\epsilon = \frac{2 L_k^2 \epsilon'}{(\tau' + L_k)^2 + x'^2}, \qquad \epsilon' = L_k \epsilon_R e^X
\end{align}
where $\epsilon_R$ in the Rindler coordinates is taken to be constant throughout.

In what follows, we consider semi-infinite subsystems on the left and right thermal radiation bath CFT$^{\rm I,II}$s coupled to a $2d$ eternal black hole on the EOW brane, and compute the entanglement entropy using both the island and the DES formula in \cref{eq_icft_is,eq_icft_des} respectively. The subsystem $A$ is described by endpoints  $M,P \equiv (\pm x'_1,\tau'_1,\epsilon')_{\rm I}$ in radiation bath CFT$_2^{\rm I}$ and $N,Q \equiv (\pm x'_1,\tau'_1,\epsilon')_{\rm II}$ in radiation bath CFT$_2^{\rm II}$ (as depicted in \cref{fig_tdni,fig_tdi,fig_tdii}). For simplicity of computations, we choose the unprimed coordinates as the starting point, where the transformations in \cref{eq_trans1} allow us to map the endpoints of the subsystems to $(\pm x_1,\tau_1,\epsilon)_{\rm I,II}$.

Based on the values of the different parameters involved, three distinct entanglement entropy phases are observed --- the no-island phase, the island phase, and the induced island phase. We obtain an exact match between the field theory and the bulk results for all three cases.


\subsection{No Island Phase}

\begin{figure}
\centering
\includegraphics[scale=0.78]{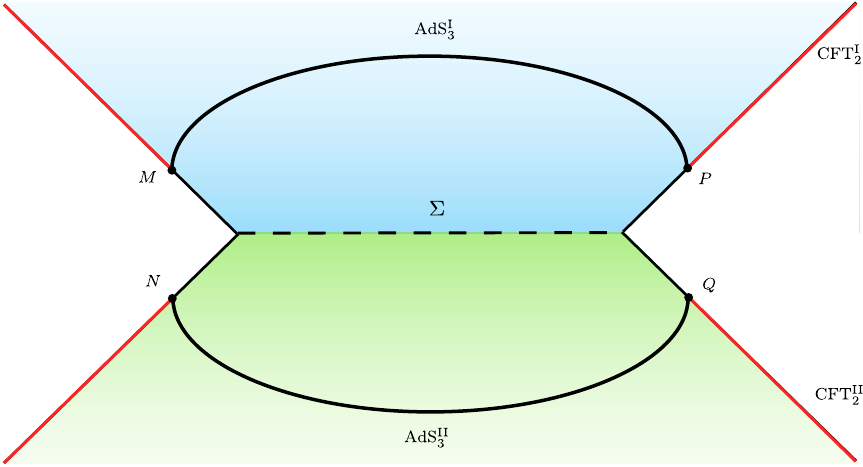}
\caption{No Island Phase in the time-dependent scenario. The radiation subsystem $A$ is given in red, while the homologous bulk RT surfaces $MP$ and $NQ$ are given by the black solid curves, which in this phase are Hartman-Maldacena type surfaces.} 
\label{fig_tdni}
\end{figure}

\paragraph{Field Theory Computations} This phase corresponds to the scenario where no entanglement entropy islands are observed on the EOW brane, as depicted in \cref{fig_tdni}. Consequently, the area term in \cref{eq_icft_is} does not contribute in this phase, and the generalized entanglement entropy (which is also the final entanglement entropy) is given in terms of the two-point function of twist fields in CFT$^{\rm I,II}$ as 
\begin{align}
S=\lim_{n \to 1} \frac{1}{1-n} \left[ \log \langle \sigma_n(M) \bar{\sigma}_n(P) \rangle_{_{\text{CFT}_2^{\rm I}}} + \log \langle \bar{\sigma}_n(N) \sigma_n(Q) \rangle_{_{\text{CFT}_2^{\rm II}}} \right].
\end{align}
The final entanglement entropy is then obtained as
\begin{align}\label{eq_td_ni_cft}
S = \frac{\cI + \cII}{3} \log \frac{2 x_1}{\epsilon} = \frac{\cI + \cII}{3} \log \frac{2 \cosh T}{\epsilon_R},
\end{align}
where we have utilized the transformations in \cref{eq_trans1,eq_trans21,eq_trans22} in the final step to express the result in the Rindler coordinates.

\paragraph{Bulk Computations} From the bulk perspective, the RT surfaces homologous to the radiation subsystem are depicted in \cref{fig_tdni} by the black curves. In this phase, they are Hartman-Maldacena (HM) type surfaces which cross the horizon from the left copy to the right copy of the thermal CFT$_2^{\rm I,II}$s. Since no corresponding island regions appear on the EOW brane, the defect term in \cref{eq_icft_des} does not contribute. Consequently, the holographic entanglement entropy in this phase is given by the sum of the lengths of the HM surfaces using \cref{eq_rt}, which matches exactly with the field theory result in \cref{eq_td_ni_cft} on implementation of the Brown-Henneaux relation $c_k = \frac{3 L_k}{2 G_N}$ \cite{Brown:1986nw}.


\subsection{Island Phase}

\begin{figure}
\centering
\includegraphics[scale=0.78]{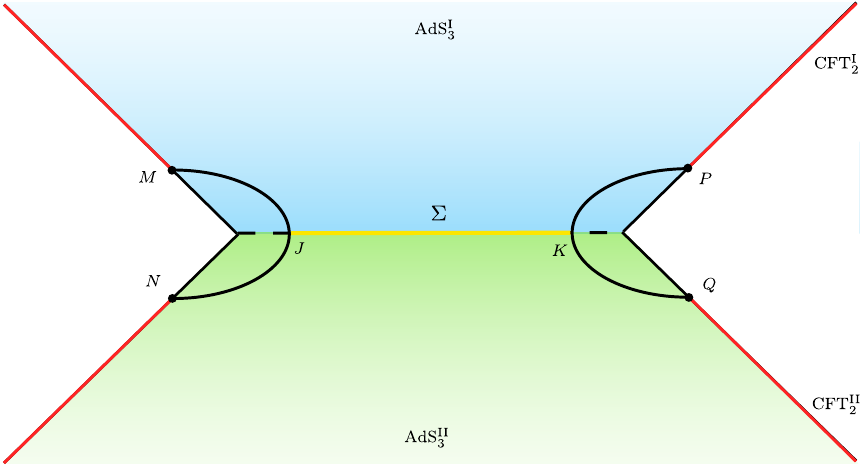}
\caption{Island Phase in the time-dependent scenario. The radiation subsystem $A$ is given in red, while the corresponding island region on the EOW brane is depicted in yellow. The homologous bulk RT surfaces $MN$ and $PQ$ are given by the black solid curves, which in this phase cross the EOW brane $\Sigma$ from one AdS geometry to the other.} 
\label{fig_tdi}
\end{figure}

\paragraph{Field Theory Computations} In this phase, we observe entanglement entropy islands on the EOW brane, which have endpoints at $J,K$ described at $(\pm x_b,y)$, with $y$ being the coordinate along the brane. This phase is illustrated in \cref{fig_tdi}. The effective entanglement entropy term in \cref{eq_icft_is} may be expressed in terms of the following four-point twist field correlators on CFT$^{\rm I}_2$, which in the large central charge limit factorize as
\begin{align}
\langle \sigma_n(M) \bar{\sigma}_n(J) \sigma_n(K) \bar{\sigma}_n(P) \rangle_{_{\text{CFT}_2^{\rm I}}} \sim \langle \sigma_n(M) \bar{\sigma}_n(J) \rangle_{_{\text{CFT}_2^{\rm I}}} \langle \sigma_n(K) \bar{\sigma}_n(P)  \rangle_{_{\text{CFT}_2^{\rm I}}},
\end{align}
with a similar factorization of twist field correlators in CFT$^{\rm II}_2$. The effective entanglement entropy term may then be obtained as the sum of 
\begin{align}
S^{\rm (I)} = \frac{\cI}{3} \left( \log \frac{(x_1-x_b)^2 + (\tau_1 + y)^2}{\epsilon \epsilon_y} + \log \frac{\LI}{y \sech \rhoI} \right),
\end{align}
and $S^{\rm (II)}$, obtained by replacing $(\cI,\LI,\rhoI) \to (\cII,\LII,\rhoII)$. The generalized entanglement entropy is then determined by adding the area term given in \cref{eq_s_area}, and is extremized at $(x_b,y)=(x_1,\tau_1)$. Putting them back into $S_{gen}$, we obtain the final entanglement entropy in this phase as 
\begin{align}\label{eq_td_i_cft}
S & = \frac{\cI}{3} \left( \rhoI + \log \frac{2 \tau_1}{\epsilon} + \log \frac{2 \LI}{\epsilon_y \sech \rhoI} \right) + \frac{\cII}{3} \left( \rhoII + \log \frac{2 \tau_1}{\epsilon} + \log \frac{2 \LII}{\epsilon_y \sech \rhoII} \right) \notag \\
& = \frac{\cI}{3} \left( \rhoI + \log \frac{2 \sinh X}{\epsilon_R} + \log \frac{2 \LI}{\epsilon_y \sech \rhoI} \right) + \frac{\cII}{3} \left( \rhoII + \log \frac{2 \sinh X}{\epsilon_R} + \log \frac{2 \LII}{\epsilon_y \sech \rhoII} \right),
\end{align}
where in the final step we have utilized the transformations in \cref{eq_trans1,eq_trans21,eq_trans22}.

\paragraph{Bulk Computations} From the bulk perspective, the RT surfaces homologous to the radiation subsystem cross the EOW brane from AdS$^{\rm I}$ into AdS$^{\rm II}$ (given by the black curves in \cref{fig_tdi}). The generalized entanglement entropy may then be obtained using \cref{eq_icft_des} as
\begin{align}
S_{gen} & = \frac{\LI}{2G_N} \log \frac{(x_1-x_b)^2+y^2+\tau_1^2+2 \tau_1 y \sinh \rhoI}{\epsilon y \sech \rhoI} \notag \\
& \qquad + \frac{\LII}{2G_N} \log \frac{(x_1-x_b)^2+y^2+\tau_1^2+2 \tau_1 y \sinh \rhoII}{\epsilon y \sech \rhoII} + S_{defect}
\end{align}
where $S_{defect}$ is given by \cref{eq_s_defect}.\footnote{\label{fn_eta} As shown in \cref{sec_TI}, the expression in \cref{eq_s_defect} corresponding to $\eta<1$ either leads to an unphysical or no solution. Consequently, in this section we directly use the expression corresponding to $\eta>1$.} $S_{gen}$ is extremized at $(x_b,y)=(x_1,\tau_1)$, and the final entanglement entropy is given as \cref{eq_td_i_cft} on implementation of the Brown-Henneaux relation.


\subsection{Induced Island Phase}

\begin{figure}
\centering
\includegraphics[scale=0.78]{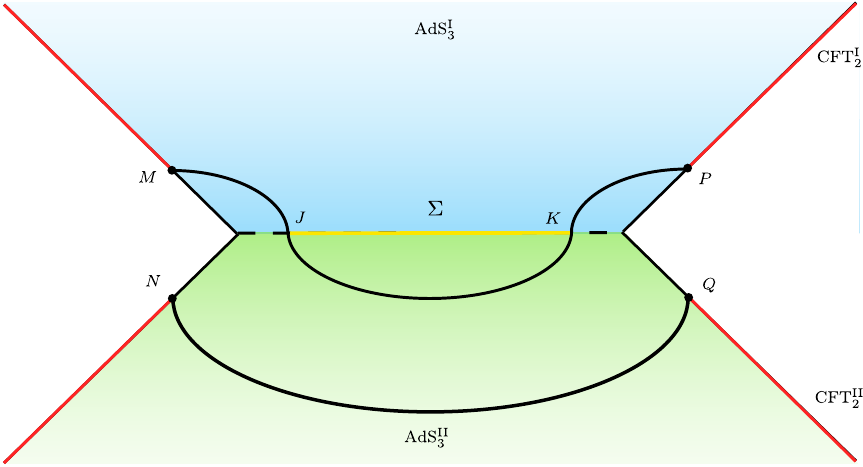}
\caption{Induced Island Phase in the time-dependent scenario. The radiation subsystem $A$ is given in red, the corresponding island region on the EOW brane is depicted in yellow, and the homologous bulk RT surfaces are given by the black solid curves. The geodesic $MP$ double-crosses the EOW brane $\Sigma$ from AdS$^{\rm I}$ to AdS$^{\rm II}$ and back. On the other hand, $NQ$ is a HM type surface in AdS$^{\rm II}$.} 
\label{fig_tdii}
\end{figure}

\paragraph{Field Theory Computations} In this phase, we observe an entanglement entropy island corresponding to the radiation subsystem in CFT$^{\rm I}$, while an induced island is observed in CFT$^{\rm II}$. As shown in \cref{fig_tdii}, the location of the islands is once again specified by $(\pm x_b,y)$ in the unprimed coordinates. The effective entanglement entropy term in \cref{eq_icft_is} may be expressed in terms of the following four-point correlators of twist fields in CFT$^{\rm I,II}$, which factorize in the large central charge limit as 
\begin{align}
\langle \sigma_n(M) \bar{\sigma}_n(J) \sigma_n(K) \bar{\sigma}_n(P) \rangle_{_{\text{CFT}_2^{\rm I}}} & \sim \langle \sigma_n(M) \bar{\sigma}_n(J) \rangle_{_{\text{CFT}_2^{\rm I}}} \langle \sigma_n(K) \bar{\sigma}_n(P)  \rangle_{_{\text{CFT}_2^{\rm I}}}, \notag \\
\langle \sigma_n(N) \bar{\sigma}_n(J) \sigma_n(K) \bar{\sigma}_n(Q) \rangle_{_{\text{CFT}_2^{\rm II}}} & \sim \langle \sigma_n(N) \bar{\sigma}_n(Q) \rangle_{_{\text{CFT}_2^{\rm II}}} \langle \sigma_n(K) \bar{\sigma}_n(J)  \rangle_{_{\text{CFT}_2^{\rm II}}},
\end{align}
The generalized entanglement entropy is then given by 
\begin{align}
S_{gen} & = \frac{\cI}{3} \left( \rhoI +\log \frac{(x_1-x_b)^2+(\tau_1 +y)^2}{\epsilon \epsilon_y} + \log \frac{\LI}{y \sech \rhoI} \right) \notag \\
& \qquad + \frac{\cII}{3} \left( \rhoII + \log \frac{2 x_1}{\epsilon} + \log \frac{2 x_b}{\epsilon_y} + \log \frac{\LII}{y \sech \rhoII} \right),
\end{align}
where we have included the area term given in \cref{eq_s_area}. $S_{gen}$ is then extremized at $(x_b,y)=(x_1,\tau_1)$ for $\cI>>\cII$\footnote{In other regimes $S_{gen}$ has unphysical solutions.}, and the final entanglement entropy is given as 
\begin{align}\label{eq_td_ii_cft}
S & = \frac{\cI}{3} \left( \rhoI +\log \frac{2 \sinh X}{\epsilon_R} + \log \frac{2 \LI}{\epsilon_y \sech \rhoI} \right) \notag \\
& \qquad + \frac{\cII}{3} \left( \rhoII + \log \frac{2 \cosh^2 T \csch X}{\epsilon_R} + \log \frac{2 \LII}{\epsilon_y \sech \rhoII} \right),
\end{align}
on implementation of the transformations in \cref{eq_trans1,eq_trans21,eq_trans22}.

\paragraph{Bulk Computations} From the bulk perspective, the RT surface homologous to the radiation subsystem crosses from AdS$^{\rm I}$ to AdS$^{\rm II}$ and back (illustrated in \cref{fig_tdii}). As a result, we observe an entanglement entropy island in AdS$^{\rm I}$ and an induced island in AdS$^{\rm II}$. The generalized entanglement entropy is then given as
\begin{align}
S_{gen} & = \frac{\LI}{2G_N} \log \frac{(x_1-x_b)^2+y^2+\tau_1^2+2 \tau_1 y \sinh \rhoI}{\epsilon y \sech \rhoI} \notag \\
& \qquad + \frac{\LII}{4G_N} \left( \cosh^{-1}  \frac{y^2 \sech^2 \rhoII+2 x_b^2}{y^2 \sech^2 \rhoII}  + 2 \log \frac{2 x_1}{\epsilon} \right) + S_{defect},
\end{align}
where $S_{defect}$ is given by \cref{eq_s_defect} (see \cref{{fn_eta}}). Once again, $S_{gen}$ is extremized at $(x_b,y)=(x_1,\tau_1)$ for $\cI>>\cII$, and the final holographic entanglement entropy is obtained as \cref{eq_td_ii_cft} on implementation of the Brown-Henneaux relation.


\subsection{Page Curve}

\begin{figure}
\centering
\includegraphics[scale=0.7]{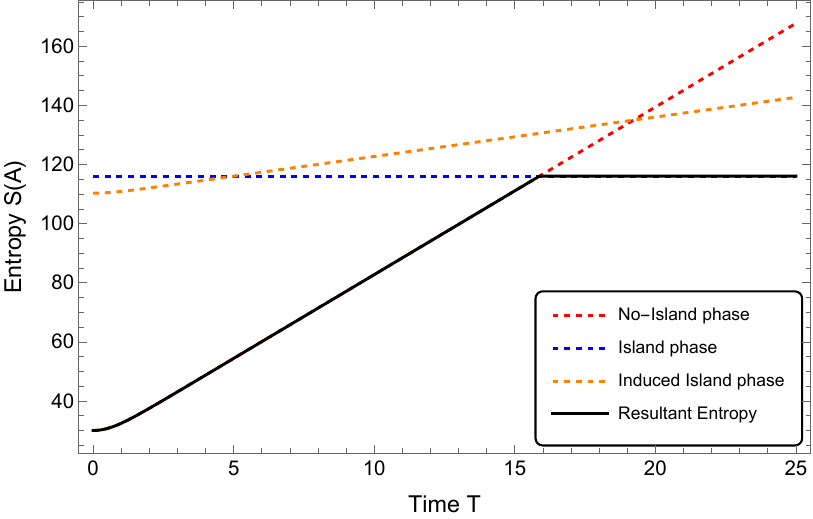}
\caption{In the Page curve given above, the colored dashed curves represent the entanglement entropy corresponding to the three phases, while the black solid curve represents the resultant entanglement entropy corresponding to the dominant phase at any given time $T$. We have set $\cI = 15, \cII = 2, \epsilon_y = \epsilon_R = 0.01, \LI = 7.5, \LII = 1, \rhoI = 2, \rhoII = 4.03, X=5$.} 
\label{fig_pcurve}
\end{figure}

We now plot the Page curve, which shows the fine-grained entanglement entropy of the radiation subsystem as a function of time. The Page curve is shown in \cref{fig_pcurve}, where the colored dashed curves represent the different entanglement entropy phases, while the black solid curve represents the minimum value of the entanglement entropy at any given time. At early times, the entanglement entropy is in the no-island phase, where it increases with time. Beyond the Page time, the entanglement entropy transitions to the island phase, where it takes a constant value. In this context, the Page time is defined as the time of transition from the no-island to the island phase and is given by
\begin{align}\label{eq_tp}
T_{\rm NI \to I} = \cosh ^{-1}\left[2 \sinh X e^{\frac{\cI \rhoI+\cII \rhoII}{\cI+\cII}} \left(\frac{\LI \cosh \rhoI}{\epsilon_y}\right)^{\frac{\cI}{\cI+\cII}} \left(\frac{\LII \cosh \rhoII}{\epsilon_y}\right)^{\frac{\cII}{\cI+\cII}}\right].
\end{align}
It is evident from \cref{fig_pcurve} that the induced island phase is never the dominant phase of the entanglement entropy as the $2d$ eternal black hole evolves in time. This may be verified analytically by computing the time of transition from the induced island to the island phase, which is given by
\begin{align}
T_{\rm II \to I} = \cosh ^{-1}\left[ \sinh X \right].
\end{align}
It can be shown that $T_{\rm II \to I} < T_{\rm NI \to I}$ for all values of the involved parameters, indicating that the entanglement entropy in the induced island phase exceeds that in the island phase even before the Page time. Thus, the induced island phase of the entanglement entropy is always subdominant throughout the time evolution of the eternal black hole.


\section{Summary and Discussions}\label{sec_summary}

In this article, we have proposed a generalization of the defect extremal surface (DES) prescription to holographic interface conformal field theories (ICFTs), thereby extending its applicability beyond the AdS/BCFT framework. The generalized prescription provides a purely geometric, higher-dimensional realization of the island formula via defect extremal surfaces that intersect at the interface brane, without invoking an explicit lower-dimensional gravitational action. We investigated the entanglement entropy for both static and dynamical settings. In the time-independent scenario, we analyzed finite bipartite subsystems in the two CFT baths. We demonstrated exact agreement between the generalized DES prescription and the effective lower-dimensional description in the large brane tension limit.

We further extended our analysis to time-dependent configurations involving semi-infinite subsystems in radiation baths coupled to $2d$ eternal black holes, where the generalized DES prescription successfully reproduces the corresponding Page curves. The Page curve exhibits the expected transition from the Hawking saddle to the island saddle, leading to the unitary saturation of the fine-grained entropy. Interestingly, although the interface geometry admits an additional induced island configuration associated with replica wormholes, we find that this phase remains subdominant throughout the parameter regime considered in this work. Consequently, the dominant contribution to the entanglement entropy is governed by the conventional island saddle, while the induced island phase does not modify the Page curve despite its geometric existence.

In both the time-independent and time-dependent scenarios, we have considered three distinct entanglement entropy phases. In addition to the conventional no-island and island phases, we include the induced island phase unique to the AdS$_3$/ICFT$_2$ framework, in which the bulk extremal surface crosses the interface EOW brane twice, passing from one AdS$_3$ geometry to the other and then returning. It is important to clarify that, although this article focuses exclusively on the double-crossing scenario, it is by no means the only saddle that permits multiple crossings. In principle, other saddles may exist in which the bulk extremal surface intersects the brane more than twice. Such phases, however, arise only in specific regions of parameter space and have been shown to possess geodesic lengths that are always larger than those considered in this work \cite{Anous:2022wqh}. Consequently, these paths do not contribute to the entanglement entropy at leading order.

The present work extends the DES prescription beyond the AdS$_3$/BCFT$_2$ framework to holographic interface conformal field theories. The resulting construction provides a higher-dimensional geometric realization of the island formula while retaining the complete braneworld geometry, without requiring an explicit lower-dimensional JT gravity description. It naturally incorporates distinctive features of holographic ICFTs, including unequal central charges, interface-crossing extremal surfaces, induced islands, and the associated replica wormhole saddles. Moreover, its exact agreement with the effective lower-dimensional description establishes the consistency of the higher-dimensional framework. Our results, therefore, broaden the scope of the DES prescription and provide a unified geometric framework for investigating quantum extremal surfaces and island phenomena in holographic interface theories, with potential applications to more general defect configurations and higher-dimensional setups.

There are several interesting directions for future investigation. An immediate extension is to apply the generalized DES prescription to holographic ICFTs deformed by irrelevant operators, particularly $T\bar{T}$ deformations, where the interaction between the finite-cutoff surface, interface EOW brane, and the bulk extremal surfaces remains unexplored. It would also be worthwhile to extend the generalized DES prescription to include mixed-state entanglement measures such as the entanglement negativity and the reflected entropy. Investigating the validity of the generalized DES framework in more general defect theories is also an interesting direction. We leave such open issues for future consideration.

\begin{appendix}

\section{The area and defect terms}\label{app_sad}

\begin{figure}
\centering
\includegraphics[scale=1]{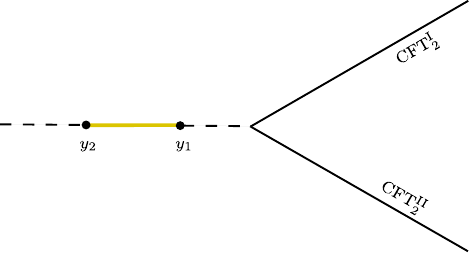}
\caption{This schematic illustrates the island region on the EOW brane.}
\label{fig_sad}
\end{figure}

Considering the island region on the EOW brane given by $I_A \equiv [y_1, y_2]$ on a constant time slice, as illustrated in \cref{fig_sad}, the area term is given by \cref{def_s_area} as
\begin{align}
S_{area} (X) = \frac{Area(\partial I_A)}{4 G_N^{(2)}}.
\end{align}
Here $Area(\partial I_A)$ is simply the area of the two endpoints of the island region on the brane in the defect AdS$_3$/ICFT$_2$ framework described in \cref{sec_desinterface}, and may therefore be set equal to 2 (the contribution for each point being 1). Finally, using \cref{eq_g2}, the area term becomes
\begin{align}
S_{area} = \frac{\cI}{3} \rhoI + \frac{\cII}{3} \rhoII,
\end{align}
where we have utilized the Brown-Henneaux relation $c_k = \frac{3 L_k}{2 G_N}$ \cite{Brown:1986nw}, and rescaled $\rho_k^0 / L_k \to \rho_k^0$ for simplicity.

On the other hand, the defect term $S_{defect}$ represents the contribution to the entanglement entropy from the bulk conformal matter localized on the brane, which in the present case is limited to the degrees of freedom confined within the island region. $S_{defect}$ is then obtained in the defect AdS$_3$/ICFT$_2$ framework as
\begin{align}
S_{defect} & = \lim_{n \to 1} \frac{1}{1-n} \left[ \log \Omega_{\rm I}(y_1)^{-2 h_n^{\rm I}} \Omega_{\rm I}(y_2)^{-2 h_n^{\rm I}} \langle \sigma(y_1) \bar{\sigma}(y_2) \rangle_{_{\text{CFT}_2^{\rm I}}} \right. \notag \\
& \qquad \qquad \qquad \qquad \left. + \log \Omega_{\rm II}(y_1)^{-2 h_n^{\rm II}} \Omega_{\rm II}(y_2)^{-2 h_n^{\rm II}} \langle \sigma(y_1) \bar{\sigma}(y_2) \rangle_{_{\text{CFT}_2^{\rm II}}} \right],
\end{align}
where the conformal factors $\Omega_k(y)$ are given in \cref{eq_cfactor}, and $h_n^k = \frac{c_k}{24}\left( n-1/n \right)$. To compute the two-point function $\langle \sigma(y_1) \bar{\sigma}(y_2) \rangle_{_{\text{CFT}_2^k}}$ on the half-plane (which in this case refers to the part EOW brane), we may consider two distinct channels corresponding to the bulk operator product expansion (OPE) and the boundary operator product expansion (BOE), respectively. The dominant channel is determined by the cross ratio $\eta = \frac{(y_2-y_1)^2}{4 y_2 y_1}$. The limit $\eta \to 0$ corresponds to a dominant OPE channel, where $\sigma(y_1)$ and $\bar{\sigma}(y_2)$ are much closer to each other than to the interface, such that
\begin{align}
\langle \sigma(y_1) \bar{\sigma}(y_2) \rangle_{_{\text{CFT}_2^k}} = \frac{\epsilon_y^{4 h_n^k}}{(y_2-y_1)^{4 h_n^k}}.
\end{align}
On the other hand, the limit $\eta \to \infty$ corresponds to a dominant BOE channel, where $\sigma(y_1)$ or $\bar{\sigma}(y_2)$ are closer to the interface than to each other. Consequently, in the BOE channel, we have
\begin{align}
\langle \sigma(y_1) \bar{\sigma}(y_2) \rangle_{_{\text{CFT}_2^k}} = \frac{\epsilon_y^{4 h_n^k}}{(4y_1y_2)^{2 h_n^k}},
\end{align}
where we have assumed that there are no additional degrees of freedom localized at the interface. Utilizing the above relations, the defect term is given by
\begin{align}
S_{defect} =
\begin{cases}
\frac{\cI}{6} \log \frac{\LI^2 (y_2-y_1)^2}{y_1 y_2 \epsilon_y^2 \sech ^2 \rhoI} + \frac{\cII}{6} \log \frac{\LII^2 (y_2-y_1)^2}{y_1 y_2 \epsilon_y^2 \sech ^2 \rhoII} \qquad & \eta <1, \\
\frac{\cI}{3} \log \frac{2 \LI}{\epsilon_y \sech \rhoI} + \frac{\cII}{3} \log \frac{2 \LII}{\epsilon_y \sech \rhoII} \qquad & \eta >1.
\end{cases}
\end{align}
The critical value $\eta_c$ for which the $S_{defect}$ transitions from one dominant channel to another may be obtained by equating the two expressions given above and solving for $y_1$ and $y_2$, which leads to $\eta_c =1$.

\section{Extremization for induced island phase}\label{app_kcurve}

\begin{figure}
\centering
\includegraphics[scale=0.8]{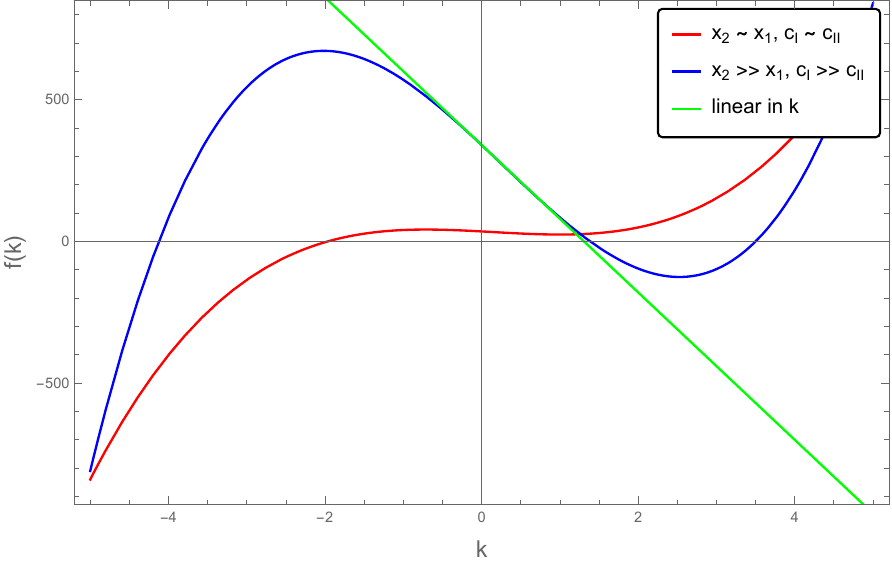}
\caption{Plot of \cref{eq_kcurve}. The red curve is for comparable $x_2 \sim x_1$ and $\cI \sim \cII$, while the blue curve is for $x_2 >> x_1$ and $\cI >> \cII$, for which one of the real positive solutions is approximately the same as that for \cref{eq_kcurve2} (given by the green line).} 
\label{fig_kcurve}
\end{figure}

The generalized entropy $S_{gen}$ for the induced island phase in \cref{ssec_tiii} is obtained by the sum of \cref{eq_tiii_s1,eq_tiii_s2} and the area term in \cref{eq_s_area}. Extremizing $S_{gen}$ for $y_1,y_2$ gives us the following conditions
\begin{align}\label{eq_con1}
\cI \frac{y_1-x_1}{y_1+x_1} + \cII \frac{y_1+y_2}{y_1-y_2} = 0, \qquad \cI \frac{y_2-x_2}{y_2+x_2} - \cII \frac{y_1+y_2}{y_1-y_2} = 0.
\end{align}
Adding the above expressions lead to the simplified condition $y_1 y_2 = x_1 x_2$, which is satisfied by the equalities 
\begin{align}\label{eq_con2}
y_1 = k x_1, \qquad y_2 = x_2/k,
\end{align}
where $k$ must have real positive values for $y_1,y_2$ to have any meaningful physical interpretation. Substituting \cref{eq_con2} in any of the expressions in \cref{eq_con1} we obtain the relation
\begin{align}\label{eq_kcurve}
f(k) = x_1 (\cI+\cII) k^3 - x_1 (\cI-\cII) k^2 - x_2 (\cI-\cII) k + x_2 (\cI+\cII) = 0.
\end{align}
It is evident from \cref{fig_kcurve} that \cref{eq_kcurve} has real and positive solutions only for $x_2 >> x_1$ and $\cI >> \cII$. In this limit, one of the solutions is approximately the same as that for the linear equation
\begin{align}\label{eq_kcurve2}
f(k) = -(\cI-\cII) k + (\cI+\cII) = 0.
\end{align}

\end{appendix}


\section*{Acknowledgments}
We would like to thank Himangshu Chourasiya, Saikat Biswas, and Ankit Anand for their feedback on the draft. The work of Ankur Dey is supported by the Fellowship for Academic and Research Excellence (FARE) at IIT Kanpur.

\bibliographystyle{utphys.bst}
\bibliography{ref}
\end{document}